\documentclass[aps,prl,groupedaddress,showpacs,amsmath,amssymb,intlimits,twocolumn]{revtex4}

\usepackage{graphicx}

\begin{document}

\title{Strong electromagnetically induced transparency in atomic media with large residual Doppler broadening}

\author{Lu\'{\i}s E. E. de Araujo}

\email{araujo@ifi.unicamp.br}

\author{Silv\^{a}nia A. Carvalho}

\author{Luciano S. Cruz}

\author{Ant\^{o}nio A. Soares}

\author{Armando Mirage$^{1}$}

\author{Daniel Pereira}

\author{Flavio C. Cruz}
\email{flavio@ifi.unicamp.br}

\affiliation{Instituto de F\'{\i}sica ``Gleb Wataghin,''
Universidade Estadual de Campinas, Campinas, SP 13083-970,
Brazil \\
$^{1}$Instituto de Pesquisas Energ\'{e}ticas e Nucleares, C. Postal 11049, 
S\~{a}o Paulo, SP 05422-970, Brazil
}

\date{\today}

\begin{abstract}

Electromagnetically induced transparency and coherent population trapping were observed in a hot (1000~K) calcium vapor embedded into an electrical gas discharge. Unexpectedly large transparencies (of up to 70\%) were observed under very unfavorable conditions: probe wavelength shorter than the coupling wavelength, and coupling Rabi frequency significantly smaller than the residual Doppler linewidth of the two photon transition. We developed a theoretical model that shows that the observed results are due to the combined effects of a strong probe beam and a small open character of the atomic system. Coherent population trapping also manifests itself as a change in the impedance of the gas discharge, and the phenomenon can be probed with high sensitivity via the optogalvanic effect.

\end{abstract}


\maketitle


In coherent population trapping (CPT), destructive interference between the different pathways to absorption of a resonant laser field can lead to a cancellation of this absorption~\cite{arimondo96}. CPT is typically studied in closed, three-level atomic systems driven by two laser fields. The driving fields prepare the atom in a coherent superposition of states decoupled from the applied fields such that the atom becomes stable against absorption from these lasers. CPT is thus characterized by a decrease in atomic fluorescence when both lasers are on resonance. In a related phenomenon, electromagnetically induced transparency (EIT) of a weak probe field is induced by a stronger coupling field resonant with a linked transition~\cite{fleischhauer05}. In EIT, the coupling field produces an Autler-Townes splitting of the common state, and destructive interference occurs in the probe's absorption whenever this splitting is smaller than the natural linewidth of the common state. An increase in the probe's transmission is then detected. Both CPT and EIT have found numerous applications over the recent years, including control of refractive index, lasers without inversion, slow light and quantum memories, nonlinear optics and transparency in the continuum (see \cite{arimondo96} and \cite{fleischhauer05} and references therein).

In hot atomic vapors, EIT results from averaging the Autler-Townes doublets from all atomic velocity classes. Good transparency of a weak probe beam can be observed if the coupling Rabi frequency is greater than the residual Doppler linewidth of the two photon transition: $|(\vec{k}_p + \vec{k}_c) \cdot \vec{u}|$, where $\vec{k}_{p,c}$ is the probe/coupling wave vector, and $\vec{u}$ is the most probable atomic velocity~\cite{banacloche95}. In a three-level cascade system, to minimize this residual Doppler linewidth and the required coupling power, the probe and coupling beams should be aligned in a counterpropagating geometry~\cite{banacloche95}. But, when the coupling wavelength $\lambda_c$ is longer than the probe wavelength $\lambda_p$, the Autler-Townes components of the moving atoms strongly overlap with the transparency for atoms at rest, obscuring the transparency~\cite{boon99}. Thus, the most unfavorable configuration for observing EIT in Doppler broadened media consists of copropagating coupling and probe beams with $\lambda_c > \lambda_p$ interacting with a cascade system. To our knowledge, EIT has never been observed in such a situation.

In this Letter, we describe the observation of strong EIT and CPT resonances in a gas of neutral calcium atoms embedded into a krypton-gas electrical discharge produced by a hollow cathode lamp (HCL). Alkaline-earth atoms, such as calcium, have been receiving increasing attention in the last years due to their applications in optical atomic clocks~\cite{degenhardt05}, ultracold collisions~\cite{nagel05}, and Bose-Einstein condensation~\cite{takasu03}. These helium-like elements lack hyperfine structure, and can offer nearly ideal approximations to truly closed two- and three-level systems. In particular, EIT was recently explored with these elements in new schemes for optical clocks~\cite{santra05} and laser cooling~\cite{dunn06}. In the cascade level system explored here, $\lambda_c = 586$~nm and $\lambda_p = 423$~nm, that is: $\lambda_c > \lambda_p$. Nevertheless, significant transparencies (of up to 70\%) of the blue (423~nm) probe beam were observed for both counter and copropagating beam configurations, even for a coupling Rabi frequency significantly smaller than the residual Doppler linewidth. In addition, we report the first nonoptical observation of CPT resonances via the optogalvanic effect. To our knowledge, the only two other nonoptical detection schemes of CPT that have been previously implemented consisted of monitoring a charge-exchange reaction signal between projectile ions and excited atoms from an atomic beam~\cite{gieler92} and an ionization signal from photoassociated cold atoms~\cite{dumke05}. But more significantly, our nonoptical detection consists in observing a change of an electrical property---the impedance---of the gas discharge caused by CPT.

The atomic system investigated is shown in Figure~1a. A probe field is scanned across the calcium blue (423~nm) $(4s^2) \, {}^1S_0  \rightarrow (4s4p) \, {}^1P_1$ transition in the presence of a coupling field resonant with the yellow (586~nm) $(4s4p) \, {}^1P_1 \rightarrow (4p^2) \, {}^1D_2$ transition. The lower transition is commonly used for laser cooling of calcium, since the intermediate $(4s4p) \, {}^1P_1$ state has a high natural decay rate of $\gamma_1 = 2\pi \times 34$~MHz. 
The uppermost $(4p^2) \, {}^1D_2$ state decays to the $(4s4p) \, {}^1P_1$ state at a rate of $\gamma_2 = 2\pi \times 11$~MHz, and also decays to the $(4s5p) \, {}^1P_1$ state (not shown in the figure) at a rate $\gamma_3 = 2\pi \times 0.18$~MHz~\cite{vaek91}. The $(4s5p) \, {}^1P_1$ state, in turn, decays to the metastable $(3d4s) \, {}^1D_2$ state, which has a lifetime much longer than the diffusion time of the atoms out of the laser beams, making this an open system.

Figure~1b shows the experimental setup. Two laser beams, both with a 3.0-mm diameter, enter the HCL either counter or copropagating with each other. In both cases, the lasers have orthogonal polarizations and spatially overlap throughout the extension of the lamp. The blue laser is obtained from a frequency-doubled, home-built Ti-Sapphire laser, which delivers up to 16~mW of blue light at the entrance to the lamp, with a linewidth of approximately 1~MHz. The yellow laser is from a Coherent 699 ring-dye laser, with up to 150~mW at 586~nm (and less than a 1~MHz linewidth) at the lamp.

\begin{figure}
\includegraphics{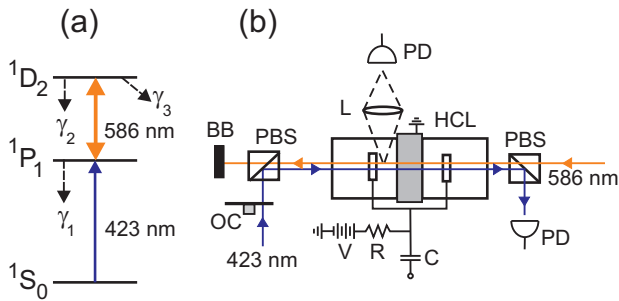} 
\caption{(Color online) (a) The open, three-level-cascade Ca system. Population may leak out of the system through the ${}^1D_2$ state. (b) The experimental setup: PD, photodetectors; PBS, polarizing beam splitters; OC, optical chopper; L, lens; V, power supply; R, ballast resistor; C, capacitor; and HCL, hollow cathode lamp. The counterpropagating beams were displaced in the figure for better viewing. For a copropagating configuration of the beams, the beam block (BB) is removed, and the coupling yellow (586 nm) beam goes in through the left beam splitter.}
\end{figure} 
The calcium atomic vapor is created by sputtering induced by a DC hollow cathode discharge (120~V, 100~mA). With no lasers present, most calcium atoms are in their ground state. Collisions with electrons or ions populate the excited states, and also ionize both sputtered and buffer gas atoms. Numerous other direct and indirect collision mechanisms also contribute to ionization. Upon laser excitation, these ionization processes are perturbed, and the impedance of the discharge is changed, generating an optogalvanic signal~\cite{barbieri90}. Our HCL is a home built lamp~\cite{cavasso01} filled with krypton gas at 2.0 torr. It is 26-cm long and made of Pyrex glass with optical windows. Two ring titanium anodes are placed 0.5~cm on each side of a 20-mm long, cylindrical steel cathode where a small calcium tube (3-mm bore) is inserted. The optical density can be easily changed by varying the lamp current, going from an optically thin to a very thick medium for the blue probe beam. At 30~mA, the Ca atomic density is on the order of $10^{10}$~cm$^{-3}$. Typical Doppler widths (full width at half maximum, FWHM) range from 1.3 to 2.2~GHz, corresponding to temperatures from 750 to 1200~K~\cite{cavasso01}. 

In the setup of Figure~1b, absorption of the probe laser beam is detected in three different ways: an optogalvanic signal is detected by measuring the voltage across the electrodes; transmission of the blue laser through the HCL is monitored by a photodetector; and fluorescence from the region in the lamp between the first (left) anode and the cathode is imaged by a lens into another photodector. The blue laser is amplitude modulated by an optical chopper at 2~kHz, and the optogalvanic, transmission and fluorescence signals are detected by three independent lock-in amplifiers. 

Figure~2a shows the detected absorption of the blue beam as a function of its detuning from the intermediate ${}^1P_1$ state. The yellow coupling beam was fixed on resonance and counterpropagating to the blue beam. The current in the lamp was set at 10~mA. The intensity in the blue and yellow lasers were 85~mW/cm$^2$ (Rabi frequency $\Omega_p \approx 0.4 \, \gamma_1$) and 707~mW/cm$^2$ ($\Omega_c \approx 1.1 \, \gamma_1$), respectively. All three absorption signals (optogalvanic, transmission, and fluorescence) consist of a wide Gaussian Doppler profile with a clear CPT/EIT resonance at zero detuning. The residual Doppler linewidth for this system is approximately $2\pi \times 245$~MHz, or $\approx 7.2 \, \gamma_1$; therefore, much larger than the coupling Rabi frequency $\Omega_c$.

An increase in the probe's transmission (EIT) is seen on resonance, corresponding to a 25\% decrease in absorption of the probe, measured as a contrast against the Doppler absorption profile in the absence of the coupling laser. The amount of induced transparency increases with the probe intensity, saturating at near 70\% at a probe intensity around 180~mW/cm$^2$, or $\Omega_p \approx 0.6 \, \gamma_1$. Both the fluorescence and transmission signals reverse direction as the probe beam reaches resonance, but the optogalvanic signal does not. In the optogalvanic signal for this particular transition, CPT is characterized by a further decrease in the voltage drop across the electrodes, and correspondingly, in the HCL's impedance.

The CPT resonances in Figure~2a have different FWHM widths: 379~MHz, 257~MHz, 233~MHz for the optogalvanic, fluorescence and transmission signals, respectively. One reason for this difference is that these signals monitor absorption over different lengths in the HCL. The transmission signal corresponds to absorption over the entire length of the lamp; the optogalvanic signal reflects the absorption across the two anodes; and the fluorescence signal gives the absorption between the first anode and the cathode. At 10~mA, the lamp is optically thick for the blue probe beam. At much lower currents, corresponding to an optically thinner medium, we observed approximately the same CPT window in the three signals. Although sub-Doppler, the measured linewidths are much larger than the natural coherence-decay rate for this system. The narrowest linewidths we measured for the transmitted signal, at much lower intensities, were about 125~MHz. 

\begin{figure}
\includegraphics{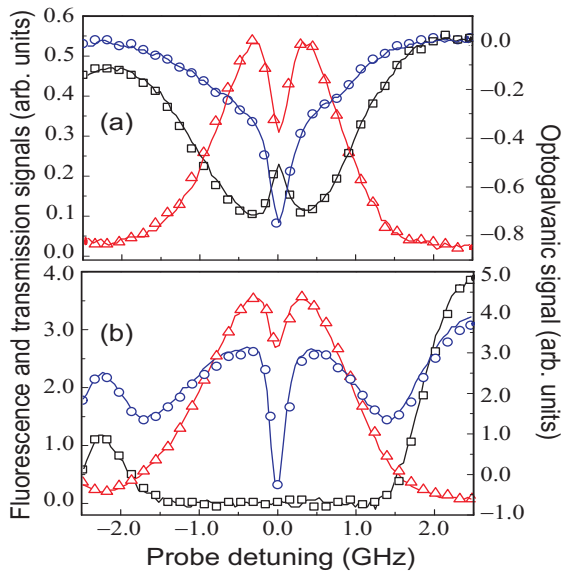}
\caption{(Color online) The optogalvanic (blue O), transmission (black $\square$) and fluorescence (red $\triangle$) signals for an HCL current of (a) 10~mA and (b) 50~mA, and counterpropagating coupling and probe beams. The estimated Rabi frequencies of the probe and coupling beams are $\Omega_p \approx 0.4 \, \gamma_1$ and $\Omega_c \approx 1.1 \, \gamma_1$, respectively, in both cases.}
\end{figure} 
Figure~2b shows the same three signals for a higher lamp current (50~mA) at which the atomic medium is optically much denser, and no light is transmitted, even in the presence of the strong coupling laser. Still, the blue laser can propagate past the first anode, and a fluorescence signal is observable. The CPT ``dip'', however, is smaller than the one in Figure~2a. At this higher current, the lineshape of the optogalvanic spectrum is changed considerably as a consequence of the atomic medium being optically very thick. Because the optogalvanic signal is a measure of the impedance change only in the region between the two anodes and the cathode, the probe laser Doppler spectrum (in the absence of the coupling laser) shows a broad flat region on its profile just because absorption is stronger at the center and weaker in the wings. Near the line center, the probe laser can be completely absorbed before reaching the electrodes, while it still reaches the region between electrodes at the wings of the Doppler profile. Therefore, this Doppler optogalvanic spectrum appears to show two side lobes at approximately $\pm 1.5$~GHz. When the coupling laser is present, a large CPT dip appears at zero detuning (Figure~2b). We observed that optogalvanic detection can be much more sensitive for detecting CPT and EIT resonances in optically thick media, when detecting probe transmission might be difficult. Optogalvanic detection might also be very convenient for detecting CPT in weak transitions, such as the ${}^1S_0  \rightarrow {}^3P_1^0$ intercombination ``clock'' transition in alkaline-earth atoms (374~Hz linewidth for Ca).

We also observed CPT in the cascade system of Figure~1 with the two beams propagating in the same direction. Figure~3 shows the observed fluorescence spectra for both counter and copropagating beam configurations. Under similar conditions of intensities and lamp current, the two configurations resulted in approximately the same cancellation of absorption at zero detuning of the probe beam.

\begin{figure}
\includegraphics{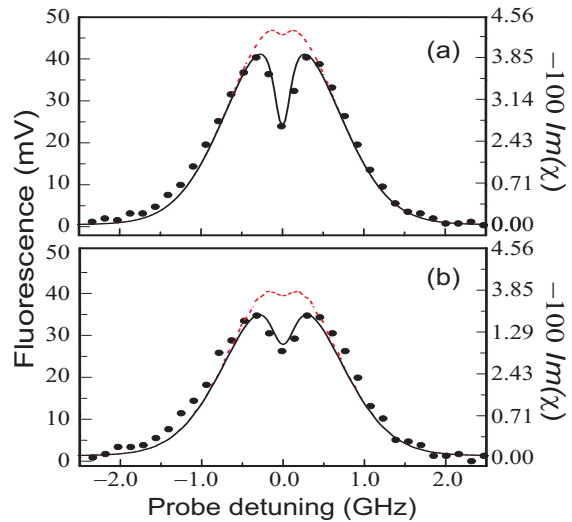} 
\caption{(Color online) The detected fluorescence signal (closed circles) and the calculated linear susceptibility $\chi$ (solid black line) of the atomic system for (a) counter and (b) copropagating beams. The red dashed line corresponds to the susceptibility $\chi$ for a closed system ($\gamma_3 = 0$).}
\end{figure}
To identify the origin of these surprisingly large CPT resonances, we numerically solved the semiclassical Bloch equations for the density matrix elements in the steady state regime. We assumed an optically thin medium, and propagation effects were not included. Velocity changing collisions were included in the model assuming the strong collision regime~\cite{quivers86}. We also included an incoherent collisional pumping rate from the ground state to the excited states. We did not measure this rate for our lamp, but up to $2\pi \times 30$~kHz, the model shows that it has no significant influence on the CPT resonances. For the laser beams employed in the experiment, the atoms have an estimated diffusion rate of approximately $2\pi \times 34$~kHz that is much faster than the decay rate of the metastable $(3d4s) \, {}^1D_2$ state: $2\pi \times 6$~Hz.

By using the known values for the transition rates and wavelengths (Figure~1), as well as the experimental Rabi frequencies, we could reproduce the experimental results without any fitting parameters. The solid lines in both Figures~3a and 3b show the imaginary part of the calculated velocity-averaged linear susceptibility $\chi$ (proportional to the absorption coefficient) of the atomic system for counter and copropagating beams, respectively. A CPT resonance is seen for both configurations, and agreement with the experiment is excellent. The linewidths of the calculated spectra are the same as those of the experiment, and they are power broadened by the strong beams. Much narrower resonances are difficult to observe since, in this wavelength mismatched system, CPT resonances are seen only with relatively strong beams. We verified that velocity-changing collisions, which are well known to limit the resolution in sub-Doppler spectroscopy in HCLs~\cite{cruz94},  have no influence on our observations and can be neglected.

It is known that saturation effects associated with a strong probe beam can significantly alter the electromagnetically induced transparency~\cite{wielandy98}. These effects, however, generally decrease the induced transparency or even enhance the probe absorption. As an example, a weak enhanced absorption was reported in an ytterbium HCL for a closed cascade system interacting with counterpropagating mismatched probe ($\lambda_p = 399$~nm) and coupling ($\lambda_c = 1077$~nm) beams with relatively high Rabi frequencies~\cite{yoon04}. Therefore, CPT and EIT resonances as large as those of Figure~2 are not expected for a closed level system interacting with strong beams, especially if they are coprapagating. If we set $\gamma_3 = 0$ in the model, and thus close our level system, the CPT resonance is strongly reduced as shown in Figure~3. We then verify that a small population decay out the atomic system dramatically affects the velocity averaged coherence created by the strong beams in a cascade system. The physical reason is that this decay decreases the amplitude of the Autler-Townes components for the each of the different atomic velocity classes that overlap with the transparency of the stationary atoms, thus increasing the overall transparency. This decrease in amplitude happens for both counter and copropagating alignment of the beams. But for a weak probe there is no significant difference between the Autler-Townes doublets of closed and open cascade systems.

From the real part of the calculated susceptibility, and for counterpropagating beams, the atomic system exhibits close to resonance a steep, linear anomalous-dispersion curve. Due to the relatively small atomic density ($\approx 10^{10} \, \text{cm}^{-3}$) in our HCL, a group index of refraction of only $n_{gr} \approx -13$ is expected. While atomic densities as high as $10^{12} \, \text{cm}^{-3}$ can be reached in an HCL, such densities are common in alkaline-metal vapor cells, and large negative index of refraction ($n_{gr} \lesssim -1500$) could be easily achieved in these media, leading to superluminal pulse propagation without significant absorption nor distortion. For copropagating beams, a nearly-flat normal dispersion is observed close to resonance.

To conclude, we observed strong CPT and EIT in a Doppler-broadened atomic medium for which the residual Doppler linewidth largely exceeded the coupling Rabi frequency. Still, under this unfavorable condition, transparencies as large as 70\% were observed. These surprising results are obtained only for significant values of the probe Rabi frequency when there is a small decay from the uppermost state. The large transparencies we observed can be used, for example, to implement optical switches in which the information content in one optical field is transferred to a second field at a much higher optical frequency. Although we discussed the case of a wavelength mismatched system, our findings are generally true for matched systems as well. Since most atomic systems are open to some degree, the results presented here could then be important to applications of CPT that employ a strong probe, such as those aiming at enhancing nonlinear optical processes~\cite{merriam00}. And for closed atomic systems, decay out of the system could be artificially induced by another weak laser. We also observed that the impedance of a gas discharge can be changed as a result of the quantum interferences associated with CPT. As such, this impedance change differs from those associated with the regular optogalvanic effect. Optogalvanic detection can be explored to study CPT involving weak transitions for which optical detection is difficult.

\begin{acknowledgments}

The authors acknowledge the financial support of FAPESP, CNPq and CAPES.
\end{acknowledgments}

\end{document}